\documentclass[mathleft
]{an}
\usepackage{graphicx}
\usepackage{natbib}
\bibpunct{(}{)}{;}{a}{}{,}
\usepackage{longtable}
\usepackage{times}
\begin{document}

\Pagespan{239}{}
\Yearpublication{2016}%
\Yearsubmission{2015}%
\Month{11}%
\Volume{337}%
\Issue{88}%

\title{Peranso - Light Curve and Period Analysis Software}

\author{E. Paunzen\inst{1}\fnmsep\thanks{Corresponding author:
  \email{epaunzen@physics.muni.cz, tonny.vanmunster@gmail.com}\newline}	
\and T. Vanmunster\inst{2}}

\institute{Department of Theoretical Physics and Astrophysics, Masaryk University,
Kotl\'a\v{r}sk\'a 2, 611\,37 Brno, Czech Republic
\and 
CBA Belgium Observatory, Walhostraat 1A, 3401 Landen, Belgium}

\received{2015}
\accepted{2015}
\publonline{later}


\abstract{A time series is a sample of observations of well-defined data points obtained through repeated measurements over 
a certain time range. The analysis of such data samples has become increasingly important not only in natural 
science but also in many other fields of research.
Peranso offers a complete set of powerful light curve and period analysis 
functions to work with large, astronomical data sets. 
Substantial attention has been given to ease-of-use and data accuracy, making it one of the most
productive time series analysis software available. In this paper, we give an introduction to Peranso and its functionality.}
\maketitle

\section{Introduction} \label{introduction}

A time series is a sequence of consecutive data points (measurements) over a time interval. Time series 
are used in almost any domain of applied science which involves temporal measurements. In astronomy, we
are normally dealing with a discrete, unequally spaced, gapped, and finite data set of an independent (time) 
and dependent (for example, magnitude) variable including a noise. From a mathematical and statistical point of
view those are the most difficult time series to deal with \citep{Chat03}.

The applications of time series analysis in astronomy are manifold going through all wavelength regions
and object classes. One of the oldest, most intriguing, and well-known problems is the analysis of the sunspot numbers and
the solar cycle \citep{Vaqu12}. The applied techniques and methods are wide-spread from very general tools
such as classical Fourier algorithms to highly specific developed tools for X-ray observations \citep{Subb97}.
The broad range of applications and techniques is nicely summarized in \citet{Feig12}.

With the availability of high precision and long term satellite based data, the search for variability of all kinds
is more efficient and successful than ever. The satellite projects CoRoT,
Kepler, and MOST resulted not only in the detection of variable stars but also of exoplanet transits \\ \citep{Zwin13,Mout14,Zasc15}. 

Several stand-alone programms and tools for the analysis of astronomical time series are available such as CINDERELLA \citep{Reeg08},
Period04 \citep{Lenz05}, and VARTOOLS \citep{Hart08}, just to mention a few of them. 
They are developed in different program languages for several operating systems. In this paper, we will review 
Peranso (developed by the second author), which provides an extensive range of functionalities for light curve and period analysis. 

In this paper, we give an introduction to the functions, modules, and tasks currently implemented in Peranso (v2.51).
The software\footnote{http://www.peranso.com} is shareware. A very extensive user manual is available for 
download\footnote{http://tonnyvanmunster.ipage.com/peranso/PeransoUserManual.pdf}.

\section{Data Input/Output} \label{data_io}

Peranso supports many different file formats, such as plain ASCII and comma delimited (CSV). The user can select to import data from a file
or from the Windows Clipboard. There is also a special option to select for the traditional Unix/Linux line-separator of text files. 
In principle, there are two basically different file formats ``free'' and ``fixed''. For both of them the user can choose to skip
a certain amount of lines. For the free format, there is a choice of how many columns are available, which information they contain and
if certain columns should be skipped. It is therefore not necessary to have a file with the independent value in the first, and the dependent
variable in the second column. The fixed format includes files of the following types:
\begin{itemize}
\item {\it AAVSO:} file formats used before and after 2009 \citep{Hend10} 
\item {\it AIP4Win:} AIP for Windows v1.4.x and v2.x \citep{Berr05}
\item {\it ASAS:} All Sky Automated Survey \citep{Pojm02}
\item {\it NSVS:} Northern Sky Variability Survey \citep{Wozn04}
\end{itemize}
All generated results can be exported as plain text files or copied to the Windows Clipboard.

\section{Time Series Analysis Methods} \label{tsam}

In this Section, an overview of the implemented time series analysis methods are given. 

\subsection{Fourier Methods} \label{Fourier}

For any kind of signal processing, the Fourier transform is the tool to connect the time domain and the frequency domain
\citep{Bloo76}. 
As a diagnostic tool, normally, the power spectrum of a time series is used. It is the square of the amplitude of each
harmonic, and it provides the contribution of each harmonic to the total energy of the time series. 

Peranso includes six different Fourier methods which we will now describe in more detail.

{\it Discrete Fourier Transform (DFT):} the DFT converts a finite list of equally spaced samples of a function (for example, a 
photometric time series) into the list of coefficients of a finite combination of complex sinusoids. The frequencies of the resulting 
sinusoids are integer multiples of a fundamental frequency, whose corresponding period is the length of the sampling interval. 
\citet{Deem75} was among the first who introduced the DFT for the analysis of astronomical unequally spaced data. 
He demonstrated that the Fourier and power spectrum analysis of such data are comparable to the analysis with equally-spaced data.
The main difference is that the aliasing has to be derived from the spectral window function.

{\it Date-Compensated Discrete Fourier Transform (DCDFT):} this technique corresponds to a maximum likelihood 
sinusoidal regression curve-fitting. It is defined such to include the uneven spacing of the time of observations and 
weighting of the corresponding data. The formalism listed by \\ \citet{Ferr81} allows to derive a better determination 
of the spectral intensity and to design harmonic filters. Furthermore, he demonstrated that the determination of the signal amplitude 
is more accurate than by DFT, when the sampling is unequally spaced, and/or there is a period longer than the total 
time base of the data set.

{\it Lomb-Scargle algorithm:} it is a variation of the DFT, in which an unequally spaced time series is decomposed into a 
linear combination of sinusoidal and cosinusoidal functions \citep{Lomb76,Scar82}. The data are transformed from the time to 
the frequency domain (Lomb-Scargle periodogram), invariant to time shifts. From a statistical point-of-view, the resulting
periodogram is related to the $\chi^{2}$ for a least-square fit of a single sinusoid to data which can treat heteroscedastic 
measurement uncertainties. The underlying model is non-linear in frequency and the basis functions at different frequencies 
are not orthogonal. Here, the modifications of \citet{Horn86} are incorporated. Basically, the power of the modified periodogram 
is normalized by the total variance of the data, yielding a better estimation of the frequency of the periodic signal.

{\it Bloomfield algorithm:} it is quite similar to the Lomb-Scargle algorithm, and also calculates a power spectrum, starting from 
unequally spaced data, using the Least Squares Standard Technique \citep{Bloo76}.

{\it CLEANest:} this method of removing false peaks from the power spectrum was introduced by \citet{Fost95}.
It applies the optimal discrete Fourier representation of the data (discrete spectrum) and the Fourier transform of the 
residuals (residual spectrum). The discrete spectrum is formed by the individual amplitudes of each frequency component 
that is used to construct the CLEANest model function. They are represented by horizontal lines in the spectrum, 
drawn at the identified frequencies, and have no width (only an amplitude). The residual spectrum is obtained by 
subtracting the model function from the original data and Fourier analyzing the residuals by a DCDFT. 
Peranso also implements the SLICK spectrum/method \citep{Fost95}, which is a very useful tool for extracting multiple signal 
components from a given data set. SLICK iteratively searches for multiple frequencies in a given signal, and attempts 
to find a ``best-fitting ensemble'' of frequencies. SLICK will adjust each found frequency such that the overall signal 
strength is maximized. Both methods are combined in one convenient Peranso dialog box, called the CLEANest Workbench.

{\it Fourier Analysis of Light Curves (FALC):} \citet{Harr89} developed this algorithm that takes multiple light 
curve segments and performs a Fourier analysis on the data. For each light curve segment, a new magnitude level (zero-point) is assumed. 
It is also possible to do a linear least squares fit for a specified period up to any harmonic order.
Furthermore, it allows different parameters such as number of harmonics, period, size of period steps, and so on to
be held constant while others are varied. This method takes the observational errors for the period determination and its
uncertainty into account. In Peranso there is a so-called ``FALC Workbench'' included which migrates the original
FALC features in a modern Windows GUI environment. It provides more sophisticated tools and outputs, for example showing the uncertainty of the fitted curve,
keeping a period constant and increment harmonic orders, and determining the most significant fit order to work with.
Within this Workbench it is also possible to derive the significance of the harmonics for a given test period.

\subsection{String-Length Methods} \label{slm}

These methods are all based on very simple assumptions. First of all, the data are folded on a series of trial periods.
For each of them, the original data are assigned phases which are then re-ordered in ascending sequence. The re-ordered data are 
examined by inspection across the full phase interval between zero and one. For each trial period 
the sum of the lengths of line segments joining successive points (the string-length) are calculated. 
Minima in the plot of the string-length versus the trial period can be considered as corresponding
to the underlying period. The methods are especially useful for a very small number of randomly spaced observations. 

The differences between the implemented methods are mainly about the treatment of observational errors, unfavourable phase
distributions, and the detection probabilities. 

{\it Lafler-Kinman algorithm:} this was the first published astronomical application of the string-length method \\ \citep{Lafl65}. 
They already give expressions for the number of trial periods needed and the number of observations required
as a function of the detectable amplitude for a known noise level to prevent spurious periods.

{\it Renson algorithm:} \citet{Rens78} improved the basic algorithm from \citet{Lafl65} by introducing a free parameter which is
sensitive to large gaps in the data set. 

{\it Jurkevich algorithm:} after all data are assigned to groups according to their phases around each trial period, the variance for each group 
and the sum of all groups are calculated \citep{Jurk71}. In addition, the variance for the complete data set can be computed. The normalization factor is derived 
by dividing the sum of all groups through the variance of the complete data set. If a trial period equals the true one, then the normalization factor 
reaches its minimum. A true period will give a much reduced variance relative to those given by other false-trial periods and with almost constant values.
In Peranso, the slightly modified form proposed by \citet{Morr80} is incorporated.

{\it Dworetsky algorithm:} more sophisticated criteria were established for predicting the expected value of the minimum string-length
in the presence of a known amount of noise in the data \citep{Dwor83}. 

{\it Phase-binned Analysis of Variance (AoV):} first the ratio of the sum of inter-bin variances to the sum of intra-bin variances 
is calculated \citep{Schw89}. The statistic follows a $F$ distribution assuming normal homoscedastic errors and a sufficiently 
large data set. This method might be preferable because the numerator and denominator are independent of each other and it is 
insensitive of the shape of the variation in the binned phase time series.

{\it Phase Dispersion Minimization (PDM):} this method might be considered as a special case of a string-length method \citep{Stel78}.
For each trial period, the full phase interval between zero and one is divided into a user defined number of bins. The width of each bin 
is also defined by the user, such that either a data point is not picked (if a bin width is selected that is 
narrower than the bin spacing), or an observation point can belong to more than one bin (if a bin width is selected that 
is wider than the bin spacing). The variance of each of these bins is then calculated, giving a measure of the scatter around the mean 
light curve, defined by the means of the data in each sample. The PDM statistic then is calculated by dividing the overall variance of 
all the samples by the variance of the original (unbinned) dataset. This process is repeated for each next trial period. 

\subsection{Method for Eclipsing Algol-type Binaries} \label{EA}

Patrick Wils (private communication) developed a novel period search method that takes advantage of photometric 
survey data, for example from the All Sky Automated Survey \citep[ASAS-3,][]{Pojm02}, of eclipsing Algol-type (EA) binaries.
The method uses only the observations that correspond with a faint state of the variable (close to the eclipse). By choosing at least three 
data points of faint states, it searches for the correct period which then can be improved using a phase binned plot.
The method might be applicable to other variable types as well.

\subsection{Analysis of Variance Methods (ANOVA)} \label{ANOVA}

The basic idea is to employ periodic orthogonal polynomials to fit the observations and the analysis of variance (ANOVA) statistic to evaluate 
the quality of the fit \\ \citep{Schw96}. The method uses multiple harmonic Fourier series as an approximation for the observed data set. 
The expansion is equivalent to an expansion into Fourier series with improvements in the detection sensitivity and in the damping of alias periods.
It employs recurrence formulae for efficient calculation of the orthogonal combinations of the harmonics, so that the number of computations scales with 
degree of freedom. It is ideally suited for the analysis of the non-sinusoidal oscillations. \citet{Schw96} showed that it might be
the best method known for the detection of non-sinusoidal oscillations with a single frequency in unevenly sampled data. 

\subsection{Edge Enhanced Box-fitting Least Squares} \label{EEBLS}

Box-fitting Least Squares (BLS) algorithms are particularly effective to analyze photometric time series in search for periodic transits by 
exoplanets \citep{Kova02}. They search for signals characterized by a periodic alternation between two discrete levels in the light curves, 
with much less time spent at the lower (fainter) magnitude level. Basically, it fits the data with a simple box function. For any given set of 
data points, the algorithm aims to find the best model with the estimators of five parameters: the period, the fractional transit length, the 
low and high states, and the epoch of the transit, respectively. For each trial period, a time series is folded to the period, which is a 
permutation of the original time series. Essentially, the method is a $\chi^{2}$ fit of a square-well transit model to the observations.
The Edge Enhanced Box-fitting Least Squares (EEBLS) is an extension to BLS, that takes into account edge effects such as limb darkening during 
exoplanet transits. Peranso allows calculating and visualizing the EEBLS frequency spectrum, folding of the time series over the most 
dominant EEBLS period, calculating the epoch of mid-transit events, the transit depth and duration, and so on. In addition, it graphically 
displays the fit obtained by the EEBLS method. 

\subsection{Planetary Transit (AoVtr)} \label{AoVtr}

This is a modification of the analysis of variance periodogram \citep{Schw89}
for the specific purpose of planetary transit search \\
\citep[AoVtr,][]{Schw06}. The underlying principle
is to assume as null hypothesis that the data are fitted by a constant value. As a next step, 
a phase-binned light curve, corresponding to a step function is employed and tested as a 
alternative hypothesis. The bin with the faintest mean magnitude is taken as the transit bin, and the 
out-of-transit mean magnitude is calculated from the remaining data points. Using the division of the data 
into out-of-transit and in-transit, the AoVtr statistic is calculated. The maximum of the AoVtr statistic over all test 
periods indicates the best transit signal for the light curve. One of the biggest advantage is that it is a one parameter fit only
(the in-transit magnitude) with a clear analytical statistical formulation and significance probability estimation.

\subsection{Spectral Window Function} \label{swf}

The spectral windows function is caused by the spectral leakage of all forms of the DFT. 
The discrete nature of the DFT yields intrinsic quantization errors and sampling artifacts (aliasing).
This implies that the frequency resolution is inversely proportional to the sample rate, for a given
number of samples and for a given sample rate, the frequency resolution is proportional to the number of
samples \citep{Koer89}. In general, the characteristics of the function is depending on the time base
of the data set, the number of data points, the distribution and lengths of the gaps as well as the
error distribution.  

The spectral window function is calculated following the method described in \citet{Deem75}.

\subsection{Period Error} \label{PE}

For the determination of the minimum period error (period uncertainty) of a given period calculating a 1$\sigma$ confidence interval, 
the method described by \citet{Schw91} is applied. It is a so-called ``post-mortem analysis'' which measures the width and height of peaks/valleys 
in a period range. Although other methods \citep{Lomb76,Scar82} are more simple and numerically faster, \\ \citet{Schw91} points out their weaknesses. Therefore
these methods were not included in Peranso. 

The applied period uncertainty method requires the estimation of the so-called Mean Noise Power Level (MNPL) in the vicinity of the test period. 
The 1$\sigma$ confidence interval of the test period is the width of the peak at the ($p$ -- MNPL) level, where $p$ is the peak height. 
Accordingly, the root mean square error ($\sigma$) is half of this confidence interval. Peranso automatically calculates an approximated MNPL value to determine 
the period uncertainty. However, low-power features could be not due to noise but due to aliasing, for example, and thus should not be taken 
into account. Therefore, it is also possible to enter a value for the MNPL manually which is then used to calculate the period error.

\subsection{False Alarm Probability (FAP)} \label{fap}

The False Alarm Probability (FAP) is a metric to express the significance of a period.
From a statistical point of view, it denotes the probability that at least one out of $N$
independent power values in a prescribed search band of a power spectrum computed
from a white-noise time series is expected to be as large as or larger than a given value.

The estimation of a statistical solid FAP is very important in order to test the probability
of a detected period. All serious time series analysis methods developed a corresponding FAP.
However, often such attempts failed in the past or are unreliable. For example, \citet{Heck85}
demonstrated that the originally proposed $F$-test of the PDM \citep{Stel78} is incorrect. The often used Lomb-Scargle 
false alarm probabilities \citep{Lomb76,Scar82} are also incorrect, because they use an improper  
equation for the number of independent frequencies \citep{Cumm99}. They also showed that the traditional Lomb-Scargle periodogram
fails when the number of observations is small, the sampling is uneven, or the period of the sinusoid is comparable
to or greater than the duration of the observations. A discussion of significance of periods for Fourier
methods can be found in \citet{Kusc97}.
 
In more practical terms, a False Alarm arises if a period is detected whereas none exists in reality.
Or in order words, the lower the FAP for a given period is, the more likely it 
is statistically significant. Normally, FAP values are expressed as a value between zero and one.
Care has to be taken in the literature because sometimes the FAP is defined as being the probability
of a null hypothesis \citep{Stur09}.

As a rule of thumb for a data set with more than 50 data points: FAPs below 0.01 (1\%) mostly indicate very secure periods, and those 
between 0.01 and 0.20 are far less certain. Anything above 0.20 (20\%) mostly relates to an 
artifact in the data, instead of a true period \citep{Horn86}.

Peranso determines the significance of a period by calculating the FAP of it using a Fisher Randomization Test \citep{Basu80}.
For this, a permutation test or Monte Carlo Permutation Procedure is performed. Those are special cases of randomization tests
that use randomly generated numbers for statistical inference \citep{Edig07}. The test executes the selected period analysis 
calculation repeatedly (at least 100 times), each time shuffling the dependent variable (for example, the magnitudes of the observation)
to form a new, randomized data set, but keeping the independent variable (for example, the observation times) fixed. 
Such a Bootstrap method is widely used in many scientific research fields \citep{Boos13}. The randomization and period calculation loop is 
performed for the number of permutations specified by the user. Potentially, a Fisher Randomization Test takes a considerable amount 
of time to execute.

Finally, two complimentary FAPs and their 1$\sigma$-errors, used for determining the significance of a test period $P$ are calculated:
\begin{itemize}
\item FAP 1 represents the proportion of permutations that contain a period with a peak higher respectively lower than the peak of $P$ 
at {\it any} frequency. It is the probability that there is no period in the investigated period range with the value of $P$.
\item FAP 2 represents the proportion of permutations that contain a period with a peak higher respectively lower than the peak of $P$ 
at {\it exactly} the frequency of $P$. It is the probability that the data set contains a period that is different from $P$.
\end{itemize}

\begin{figure}
\begin{center}
\includegraphics[width=1.0\textwidth,natwidth=850,natheight=640]{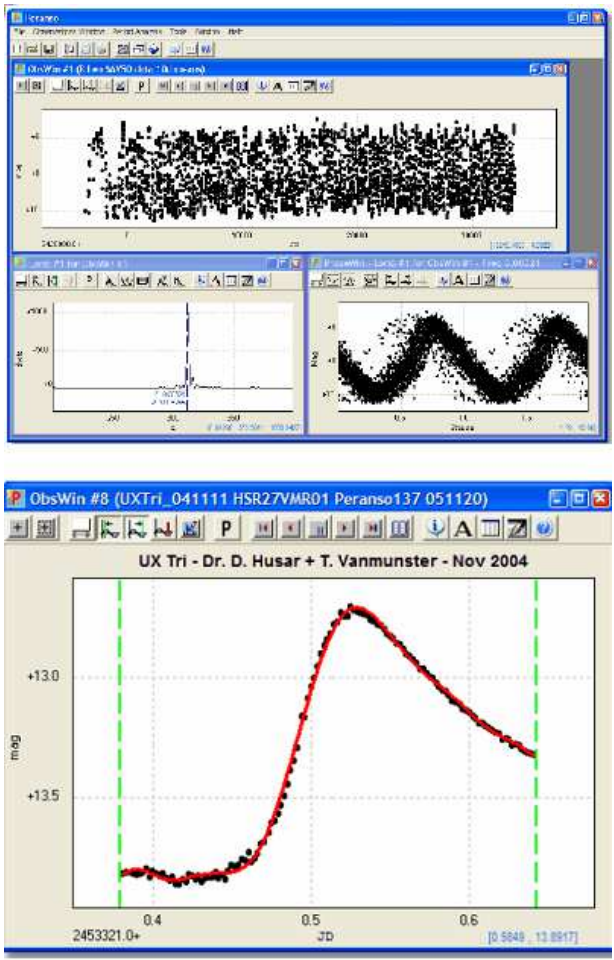}
\caption[]{Two screenshots of Peranso.} 
\label{screens}
\end{center}
\end{figure}

\section{Light curve Analysis Methods} \label{lam}

There are several modules and tasks available in Peranso to help analysing the light curves and to derive additional information.

\subsection{Finding Extrema} \label{fe}

The very popular and often used method by \citet{Kwee56} is implemented in Peranso to find the times
of extrema (minimum or maximum) in light curves. This is especially useful for minima of eclipsing binaries, for example.
This technique has two main advantages: it yields indisputable results, and works best on light curves with a symmetrical shape. It is optimal if the search interval 
consists of points that are relatively close to the extremum. 

First, the data has to be interpolated to get an equally spaced data set. This can be done in Peranso 
either via a linear or a spline interpolation \citep{Fors77}. Within the chosen interval, a starting
point for the extremum (normally the point with the brightest or faintest magnitude) is automatically chosen.
The algorithm then searched for the more appropriate extremum value using a simple quadratic formula.
This parabolic representation finally gives the time of the extremum and its mean error.  

\subsection{Auxiliary Modules} \label{aux}

There are some additional auxiliary modules available which we will briefly summarize (alphabetically) in the following.
\begin{itemize}
\item {\it Detrend All:} a linear fit (trend) can be subtracted from the data set
\item {\it Exoplanet Diagnostic:} \citet{Ting05} derived a method to detect exoplanet transits relying only on the duration, depth, and period. They showed that
it requires much less precise photometry than other methods. Basically, they derive a parameter $\eta$ called ``exoplanet diagnostic''
which is able to select transit candidates. They also show that many eclipsing binary stars, especially those involving giants, and 
blends can be immediately excluded from further consideration. The final equation for $\eta$ as included in Peranso \citep[Equ. 11 in ][]{Ting05} 
is based on the observed transit duration, the orbital parameters of the system and the stellar (planetary) radii of the components. 
The orbital and stellar parameters are very well known for classical eclipsing binaries and can be plotted versus the transit duration.
From these plots, a value of $\eta$\,$>$\,1.2 gives a high probability for being a ``planet-like'' transit. 
\item {\it Heliocentric Correct All:} a heliocentric correction can be applied if the object's right ascension and declination (J2000.0) is known. 
\item {\it Julian Day Calculator:} the calendar date can be transformed in the julian date and vice versa.
\item {\it Lightcurve Workbench:} it includes some very useful tools to easily manipulate and analyse the data. The data can be binned
in time units or number of points (observations). Furthermore, a polynomial fit up to the 50th degree can be performed and extrema 
calculated.
\item {\it Prewhitening:} the data set can be prewhitened with any period and the residuals used for a further analysis.
\item {\it Show/Hide Trend:} a linear fit (trend) can be plotted over the data set
\item {\it Time/Mag Offset:} applies a time or magnitude offset (displacement) to the data set. This can be done manually by entering a
number or graphically using the mouse.
\end{itemize}
All these features are accessible via the Peranso user interface (GUI). 

\section{The Peranso User Interface} \label{pui}

The Peranso user interface comprises basic ``Peranso window types'' and some specific graphical elements such as cursors, indicators, and
so on, together with common Microsoft Windows entities (dialog boxes, menus, toolbars, and so on). Two screenshots of Peranso are shown
in fig. \ref{screens}.

There are three basic Peranso window types
\begin{itemize}
\item The Observations Window (ObsWin)
\item The Period Window (PerWin)
\item The Phase Window (PhaseWin)
\end{itemize}
which will be discussed in more detail in the following.

{\it ObsWin:} is used for plotting and manipulating the time-series. The abscissa displays the time over which the data are plotted, 
while the ordinate represents the magnitude (or intensity). Each observation is defined by following attributes:
\begin{itemize}
\item Time 
\item Magnitude
\item Magnitude Error (MagError) [optional]: the error in the magnitude estimate. The value
is visually represented as a ``vertical bar'' ($pm$ the amount) centered around the corresponding magnitude dot in
the light curve. The MagError values are taken into account when performing a period analysis calculation using the FALC method.
\item Use status [optional]: has a value of 0 or 1 and determines if an observation is considered to
be active (1) or inactive (0). Inactive observations are not taken into account when performing
a period analysis calculation. Observations can be made active and inactive at every moment,
using the mouse and keyboard. An active observation is plotted as a filled circle whereas inactive observations 
appear as open circles.
\end{itemize}
Observations are logically grouped in observation sets. Almost all graphical properties of an ObsWin can be modified by the user.

{\it PerWin:} is used for plotting the results of a period analysis, and for
doing extensive period analysis work. The abscissa displays the time or
frequency range over which the period calculations are made. The choice between time domain or
frequency domain calculations is made at the start of a period analysis calculation. 
The ordinate displays the calculated statistic of the selected period
analysis method, or the power spectral density
\begin{itemize}
\item if a statistical method is used for the period analysis, then the ordinate displays the calculated
statistic of the selected period analysis method. For example, in the PDM method, the calculated statistic is
the PDM ``theta'' statistic.
\item if a Fourier method is used for the period analysis, then the ordinate normally displays the power
spectral density values.
\end{itemize}

{\it PhaseWin:} is used for plotting a phase diagram or folded light
curve of object's magnitude versus its phase (typically between 0 and 1). In Peranso, the JD of the very first 
observation is taken as the default epoch value, but can be adjusted by the user.

\begin{acknowledgements}
This project is financed by the SoMoPro II programme (3SGA5916). The research leading
to these results has acquired a financial grant from the People Programme
(Marie Curie action) of the Seventh Framework Programme of EU according to the REA Grant
Agreement No. 291782. The research is further co-financed by the South-Moravian Region. 
It was also supported by the grant 7AMB14AT015 and
the financial contributions of the Austrian Agency for International 
Cooperation in Education and Research (BG-03/2013 and CZ-09/2014). 
This work reflects only the authors' views, and the European 
Union is not liable for any use that may be made of the information contained therein.
\end{acknowledgements}


\begin{thebibliography}{}
\bibitem[\protect\citeauthoryear{Basu}{1980}]{Basu80} Basu, D. 1980, Journal of the American Statistical Association, 75, 575
\bibitem[\protect\citeauthoryear{Berry \& Burnell}{2005}]{Berr05} Berry, R., \& Burnell, J. 2005, The Handbook of Astronomical Image Processing, 2nd edition (Willmann-Bell, Richmond)
\bibitem[\protect\citeauthoryear{Bloomfield}{1976}]{Bloo76} Bloomfield, P. 1976, Fourier Analysis of Time Series: An Introduction (Wiley, New York)
\bibitem[\protect\citeauthoryear{Boos \& Stefanski}{2013}]{Boos13} Boos, D. D., \& Stefanski, L. A. 2013, Essential Statistical Inference: Theory and Methods (Springer Verlag, New York)
\bibitem[\protect\citeauthoryear{Chatfield}{2003}]{Chat03} Chatfield, C. 2003, The Analysis of Time Series: An Introduction, Sixth Edition (Chapman and Hall/CRC, London)
\bibitem[\protect\citeauthoryear{Cumming et al.}{1999}]{Cumm99} Cumming, A., Marcy, G. W., Butler, R. P. 1999, ApJ, 526, 890
\bibitem[\protect\citeauthoryear{Deeming}{1975}]{Deem75} Deeming, T. J. 1975, Ap\&SS, 36, 137
\bibitem[\protect\citeauthoryear{Dworetsky}{1983}]{Dwor83} Dworetsky, M. M. 1983, MNRAS, 203, 917
\bibitem[\protect\citeauthoryear{Edgington \& Onghena}{2007}]{Edig07} Edgington, E. S., \& Onghena, P. 2007, Randomization Tests, Fourth Edition (Chapman and Hall/CRC, London)
\bibitem[\protect\citeauthoryear{Feigelson \& Babu}{2012}]{Feig12} Feigelson, E. D., \& Babu, G. J. 2012, Modern Statistical Methods for Astronomy With R 
Applications (Cambridge University Press, Cambridge)
\bibitem[\protect\citeauthoryear{Ferraz-Mello}{1981}]{Ferr81} Ferraz-Mello, S. 1981, AJ, 86, 619
\bibitem[\protect\citeauthoryear{Forsythe et al.}{1977}]{Fors77} Forsythe, G. E., Malcolm, M. A., \& Moler, C. B. 1977, Computer Methods for Mathematical Computations (Prentice-Hall, New Jersey)
\bibitem[\protect\citeauthoryear{Foster}{1995}]{Fost95} Foster, G. 1995, AJ, 109, 1889
\bibitem[\protect\citeauthoryear{Harris et al.}{1989}]{Harr89} Harris, A. W., Young, J. W., Bowell, E., et al. 1989, Icarus, 77, 171 
\bibitem[\protect\citeauthoryear{Hartman et al.}{2008}]{Hart08} Hartman, J. D., Gaudi, B. S., Holman, M. J., McLeod, B. A., Stanek, K. Z., Barranco, J. A., Pinsonneault, M. H., \& Kalirai, J. S.
2008, ApJ, 675, 1254 
\bibitem[\protect\citeauthoryear{Heck et al.}{1985}]{Heck85} Heck, J., Manfroid, A., \& Mersch, G. 1985, A\&AS, 59
\bibitem[\protect\citeauthoryear{Henden}{2010}]{Hend10} Henden, A. A. 2010, The Journal of the American Association of Variable Star Observers, 38, 236
\bibitem[\protect\citeauthoryear{Horne \& Baliunas}{1986}]{Horn86} Horne, J. H., \& Baliunas, S. L. 1986, ApJ, 302, 757
\bibitem[\protect\citeauthoryear{Jurkevich}{1971}]{Jurk71} Jurkevich, I. 1971, Ap\&SS, 13, 154 
\bibitem[\protect\citeauthoryear{K{\"o}rner}{1989}]{Koer89} K{\"o}rner, T. W. 1989, Fourier Analysis (Cambridge University Press, Cambridge) 
\bibitem[\protect\citeauthoryear{Kov{\'a}cs et al.}{2002}]{Kova02} Kov{\'a}cs, G., Zucker, S., \& Mazeh, T. 2002, A\&A, 391, 369
\bibitem[\protect\citeauthoryear{Kuschnig et al.}{1997}]{Kusc97} Kuschnig, R., Weiss, W. W., Gruber, R., Bely, P. Y., \& Jenkner, H. 1997, A\&A, 328, 544
\bibitem[\protect\citeauthoryear{Kwee \& van Woerden}{1956}]{Kwee56} Kwee, K. K., \& van Woerden, H. 1956, Bulletin of the Astronomical Institutes of the Netherlands, 12, 327
\bibitem[\protect\citeauthoryear{Lafler \& Kinman}{1965}]{Lafl65} Lafler J., \& Kinman, T. D. 1965, ApJS, 11, 216
\bibitem[\protect\citeauthoryear{Lenz \& Breger}{2005}]{Lenz05} Lenz, P., \& Breger, M. 2005, Communications in Asteroseismology, 146, 53
\bibitem[\protect\citeauthoryear{Lomb}{1976}]{Lomb76} Lomb, N. R. 1976, Ap\&SS, 39, 447
\bibitem[\protect\citeauthoryear{Morris \& DuPuy}{1980}]{Morr80} Morris, S., \& DuPuy, D. L. 1980, PASP, 92, 303
\bibitem[\protect\citeauthoryear{Moutou et al.}{2014}]{Mout14} Moutou, C., Almenara, J. M., D{\'i}az, R. F., et al. 2014, MNRAS, 444, 2783
\bibitem[\protect\citeauthoryear{Pojma{\'n}ski}{2002}]{Pojm02} Pojma{\'n}ski, G. 2002, Acta Astron., 52, 397
\bibitem[\protect\citeauthoryear{Reegen et al.}{2008}]{Reeg08} Reegen, P., Gruberbauer, M., Schneider, L., \& Weiss, W. W. 2008, A\&A, 484, 601
\bibitem[\protect\citeauthoryear{Renson}{1978}]{Rens78} Renson, P. 1978, A\&A, 63, 125
\bibitem[\protect\citeauthoryear{Scargle}{1982}]{Scar82} Scargle, J. D. 1982, ApJ, 263, 835
\bibitem[\protect\citeauthoryear{Schwarzenberg-Czerny}{1989}]{Schw89} Schwarzenberg-Czerny, A. 1989, MNRAS, 241, 153
\bibitem[\protect\citeauthoryear{Schwarzenberg-Czerny}{1991}]{Schw91} Schwarzenberg-Czerny, A. 1991, MNRAS, 253, 198
\bibitem[\protect\citeauthoryear{Schwarzenberg-Czerny}{1996}]{Schw96} Schwarzenberg-Czerny, A. 1996, ApJ, 460, L107
\bibitem[\protect\citeauthoryear{Schwarzenberg-Czerny \& Beaulieu}{2006}]{Schw06} Schwarzenberg-Czerny, A., \& Beaulieu, J.-Ph. 2006, MNRAS, 365, 165
\bibitem[\protect\citeauthoryear{Subba Rao et al.}{1997}]{Subb97} Subba Rao, T., Priestly, M. B., \& Lessi, O. 1997, Applications of Time 
Series Analysis in Astronomy and Meteorology (Chapman and Hall/CRC, London)
\bibitem[\protect\citeauthoryear{Stellingwerf}{1978}]{Stel78} Stellingwerf, R. F. 1978, ApJ, 224, 953
\bibitem[\protect\citeauthoryear{Sturrock \& Scargle}{2009}]{Stur09} Sturrock, P. A., \& Scargle, J. D. 2009, ApJ, 706, 393
\bibitem[\protect\citeauthoryear{Tingley \& Sackett}{2005}]{Ting05} Tingley, B., \& Sackett, P. D. 2005, ApJ, 627, 1011
\bibitem[\protect\citeauthoryear{Vaquero}{2012}]{Vaqu12} Vaquero, J. M. 2012, in Comparative Magnetic Minima: Characterizing quiet times in the Sun and Stars, 
IAU Symposium, 286, 383
\bibitem[\protect\citeauthoryear{Wozniak et al.}{2004}]{Wozn04} Wozniak, P. R., Vestrand, W. T., Akerlof, C. W., et al. 2004, AJ, 127, 2436
\bibitem[\protect\citeauthoryear{Zasche et al.}{2015}]{Zasc15} Zasche, P., Wolf, M., Kuc{\'a}kov{\'a}, H., Vrastil, J., Jurysek, J., Masek, M., \& Jel{\'i}nek, M.
2015, AJ, 149, 197
\bibitem[\protect\citeauthoryear{Zwintz et al.}{2013}]{Zwin13} Zwintz, K., Fossati, L., Guenther, D. B., et al. 2013, A\&A, 552, A68
\end{thebibliography}
\end{document}